\documentclass[%
pre,twocolumn,
 amsmath,amssymb,
 aps,
floatfix,
]{revtex4-2}

\usepackage{graphicx}
\usepackage{dcolumn}
\usepackage{bm}
\usepackage{xcolor}
\usepackage{xr}
\usepackage{adjustbox}
\usepackage{hyperref}
\usepackage[mathlines]{lineno}
\hypersetup{
    colorlinks=true,
    linkcolor=blue,
    filecolor=magenta,      
    urlcolor=blue,
    citecolor=blue,
}
\usepackage{color}
\usepackage{listings}
\usepackage{subfigure}
\usepackage{amsthm}
\theoremstyle{definition}

\usepackage{float}  
\usepackage{tikz}
\usepackage{amsmath}
\usetikzlibrary{angles, quotes, calc}
\usepackage[normalem]{ulem}

\usepackage{lineno}

\usepackage{amsmath}
\usepackage{mathtools}

\usepackage[makeroom]{cancel}

\usepackage{verbatim}

\begin{document}

\title{Theory of Cell Body Lensing and Phototaxis Sign Reversal\\ in ``Eyeless" Mutants of 
{\it Chlamydomonas}}

\author{Sumit Kumar Birwa}
\email[]{skb61@cam.ac.uk}
\affiliation{Department of Applied Mathematics and Theoretical 
Physics, Centre for Mathematical Sciences,\\ University of Cambridge, Wilberforce Road, Cambridge CB3 0WA, 
United Kingdom}%
\author{Ming Yang}
\email[]{my365@cam.ac.uk}
\affiliation{Department of Applied Mathematics and Theoretical 
Physics, Centre for Mathematical Sciences,\\ University of Cambridge, Wilberforce Road, Cambridge CB3 0WA, 
United Kingdom}
\author{Adriana I. Pesci}
\email[]{A.I.Pesci@damtp.cam.ac.uk}
\affiliation{Department of Applied Mathematics and Theoretical 
Physics, Centre for Mathematical Sciences,\\ University of Cambridge, Wilberforce Road, Cambridge CB3 0WA, 
United Kingdom}%
\author{Raymond E. Goldstein}
\email[]{R.E.Goldstein@damtp.cam.ac.uk}
\affiliation{Department of Applied Mathematics and Theoretical 
Physics, Centre for Mathematical Sciences,\\ University of Cambridge, Wilberforce Road, Cambridge CB3 0WA, 
United Kingdom}%
\date{\today}

\begin{abstract}
Phototaxis of many species of green algae relies upon 
directional sensitivity of their membrane-bound 
photoreceptors, which arises from 
the presence of a pigmented ``eyespot" behind them 
that blocks light passing through the 
cell body 
from reaching the photoreceptor.  A decade ago it was 
discovered that the spherical cell body of 
the alga {\it Chlamydomonas reinhardtii} acts 
as a lens to concentrate incoming light, and that in
``eyeless" mutants of \textit{Chlamydomonas} the 
consequence of that focused light reaching the 
photoreceptor from behind is a reversal in the sign of phototaxis
relative to the wild type behavior.  
We present a quantitative theory of this sign reversal
by completing a recent simplified analysis of lensing [Yang, et al., \textit{Phys. Rev. E} {\bf 113}, 022401 (2026)]
and incorporating it into an adaptive model for \textit{Chlamydomonas} phototaxis. 
This model shows that phototactic dynamics in the presence of lensing 
is subtle because of the existence of internal light caustics when 
the cellular index of refraction exceeds that of water.  During each period of 
cellular rotation about its body-fixed axis, the photoreceptor receives 
two competing signals:
a relatively long, slowly-varying signal from the direct illumination, 
and a stronger, shorter, rapidly-varying lensed signal. The reversal of the sign of
phototaxis is then a consequence of the dominance of the flagellar photoresponse to the 
signal with the higher time derivative. These
features lead to a quantitative understanding of phototaxis sign reversal, 
including bistability in the direction choice, a prediction that
can be tested in single-cell tracking studies of mutant phototaxis.

\end{abstract}
\maketitle

\section{Introduction}
\label{intro}

Phototaxis exhibited by many species of green algae serves as a paradigm for the general problem of 
directional sensing by simple uni- and multicellular organisms \cite{Jekely2009}.  Unlike the gradient climbing dynamics 
found, for example, in bacterial chemotaxis \cite{Berg1976}, motion of aneural protists in 
response to light signals is generally understood to be based on a ``line-of-sight" mechanism in which light
falling on a membrane-bound photoreceptor just below the outer cell wall triggers changes in the beating dynamics
of flagella that turn the cell toward the light \cite{Schaller1997}.  
Phototaxis can be of either sign, depending on light levels and the presence of various chemical species 
in the surrounding medium \cite{Nakajima}.  For biflagellated unicellular organisms such as \textit{Chlamydomonas reinhardtii} (CR), the two flagella respond 
in opposite ways to the internal chemical changes triggered by varying illumination of the photoreceptor \cite{RufferNultsch1991}.  
As evidenced by the phototactic inability of the mutant
\textit{ptx1} which has symmetrical flagellar responses to 
light \cite{RufferNultsch1998}, this asymmetry is required for accurate phototaxis. 

\begin{figure}[b]
    \includegraphics[width=\columnwidth]{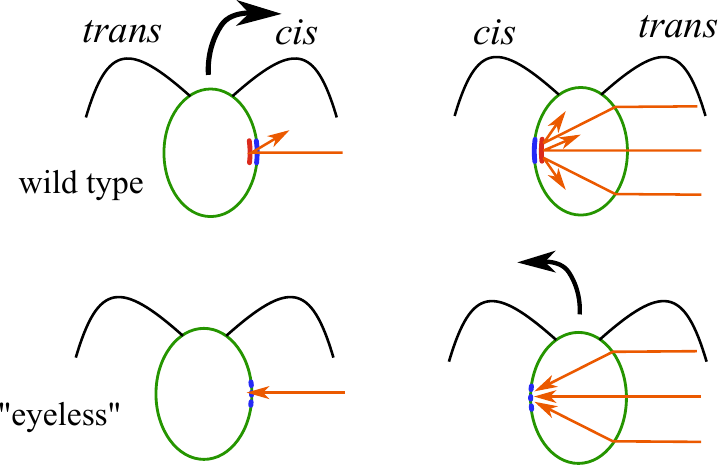}
    \caption{Phototaxis of \textit{Chlamydomonas}. Photoreceptor is shown in blue, eyespot in red.  Adapted from \cite{Ueki2016} and \cite{AlgalOptics}.  }
    \label{fig1}
\end{figure}

A long series of studies \cite{RufferNultsch1990,RufferNultsch1991,Schaller1997,Bennett2015,ChlamyPRE} 
has firmly established that CR phototaxis along with that of its multicellular relatives 
in the volvocine lineage\textemdash  
the $16$-cell \textit{Gonium} \cite{GoniumPRE} and the $1000-2000$ cell \textit{Volvox} \cite{Drescher2010}\textemdash 
can be understood quantitatively as a consequence of the interplay of two universal features:
persistent rotational motion of the organisms about a body-fixed axis and the directionally sensitivity of the photoreceptor.
The spinning motion arises from particular broken symmetries 
in the beat pattern of the flagella, for example a slight out-of-plane breaststroke beating of the CR flagella 
\cite{CorteseWan} or a tilt of the beat plane relative to the anterior-posterior axis of \textit{Volvox}.  
The directionality of the photoreception arises from the presence of a pigmented ``eyespot" behind the photoreceptor, a 
structure composed of carotenoid globules 
\cite{Seth2022} that is thought to act like a quarter-wave plate due to its thickness relative to the 
wavelength of light \cite{Foster1980} so as to reflect light 
(by constructive interference) as shown in Fig. \ref{fig1}. In this view, light 
incident on the front of the cell from the surrounding 
water would pass once through the photoreceptor, be 
reflected by the eyespot and pass through the photoreceptor 
a second time, increasing the signal in a manner analogous
to the way in which the \textit{tapetum lucidum} behind
the retina of animals such as dogs, cats and fish \cite{tapetum} 
retroreflects light. Likewise, light passing 
through the cell body from behind is blocked from reaching the 
receptor by the same reflective action of the eyespot.  

The role of the eyespot in shielding the photoreceptor 
has been the subject of debate, with some evidence
\cite{SchallerUhl1997} indicating a rather narrow wavelength-dependent peak in 
reflectivity, favoring the alternate possibility that receptor directionality arises from 
the absorptive properties of chlorophyll within the cell body.  An earlier study 
\cite{Sineshchekov1994} of mutants 
lacking chlorophyll and the eyespot investigated the 
effect of reconstituting photoreceptors with externally added retinal.  Strikingly, the sign of phototaxis observed was opposite to that of the wild type, a result suggesting that lensing by the cell body might concentrate light incident from behind the cell.

Further evidence for the possibility of cell body lensing was presented in the important
work by Kessler, et al. \cite{Kessler2015}, 
followed by the key work of Ueki, et al. 
\cite{Ueki2016} that not only demonstrated the
focusing effect directly, with an estimate of 
$\approx\! 1.47$ for the effective index of refraction $n_c$ of the cell body, but also showed that a 
number of eyeless mutants had reversed-sign phototaxis relative to the wild type.  These results 
strongly support the role of the eyespot in blocking light from behind the photoreceptor.

Motivated by cell body lensing effects in simple 
geometries, recent work \cite{AlgalOptics} has considered the general problem of 
light refraction by complex shapes found in the world of protists. It 
also used geometric optics to estimate the intensity boost $\eta$
associated with lensing of light,
defined as the ratio of light falling on the photoreceptor due to lensing relative to 
that incident from in front of the cell.
For the special case of light coming 
from directly behind a spherical cell, a considerable boost of $\eta\approx 1.5$
was found.
 
It follows from this result that as a swimming 
eyeless mutant spins around its axis, instead of just receiving 
the single pulse-like half-wave-rectified-sinusoid signal of the wild type, it receives 
a second, shorter but more intense signal during each full rotation.  
It is thus not at all clear how eyeless cells make a decision as to which signal to follow.
Do they simply follow the instantaneously stronger signal? Is there a threshold of amplification due to 
lensing for the opposite sign of phototaxis to occur?  How precise is the sign-reversed 
phototaxis?
These questions are examples of the more general issue of decision-making by aneural organisms 
confronted by competing stimuli, and have been 
recently studied in detail for CR 
responding to two separate beams of light with adjustable amplitude and angular separation
\cite{Raikwar2025}.

In this paper we present a theoretical analysis of the 
phototactic response of eyeless mutants by generalizing
the earlier optical analysis \cite{AlgalOptics} 
mentioned above to include off-axis lensing effects 
and then incorporating those results into 
the adaptive model \cite{ChlamyPRE} of CR phototaxis.
The lensing of incoming parallel light within a spherical refractive cell consists of an illuminated 
cone, outside of which is dark (see Fig. \ref{fig3}(b) below).  The outer cone boundary is a caustic, with a 
well-known peak in intensity.  This light pattern within the cell must be taken into 
account to determine the phototactic steering of a cell.  We show that even in the simplest 
treatment of these features there is competition in phototactic direction due to the 
two signals; cells can exhibit wild-type or sign-reversed phototaxis depending on 
their initial orientation with respect to the direction of the incident light.  

\section{Lensing}
\label{lensing}

A quantitative theory of phototaxis requires knowledge 
of the light intensity falling on the photoreceptor as 
a function of incident angle.  The photoreceptor occupies a roughly circular patch of 
the cell membrane, with a radius $a\sim 0.4-0.7\,\mu$m \cite{Foster1980,Kreimer2009} 
as shown in Fig. \ref{fig1} that is considerably smaller than the cell radius 
$R\approx 5\,\mu$m, and thus the receptor may be considered a planar disk.  Furthermore, 
the smallness
of the ratio 
\begin{equation}
    \epsilon=\frac{a}{R}\sim 0.08-0.14
    \label{epsdefine}
\end{equation} 
justifies the limit
$\epsilon\to 0$ adopted in calculations below.

We assume the cell is spherical and has a uniform index of refraction 
$n_{\text{c}}\ge n_w$, where $n_{\text{w}}\simeq 1.33$ is the index of refraction of the surrounding aqueous medium.
The experimental results of Ueki, et al. \cite{Ueki2016} suggested that the internal index is $n_{\text{c}} \simeq 1.47$,
which indicates the relative refractive index $n = n_{\text{c}}/n_{\text{w}} \simeq 1.1$, a representative value
we use in numerical studies of phototaxis.

\begin{figure}[t]
    \centering
    \includegraphics[width=\columnwidth]{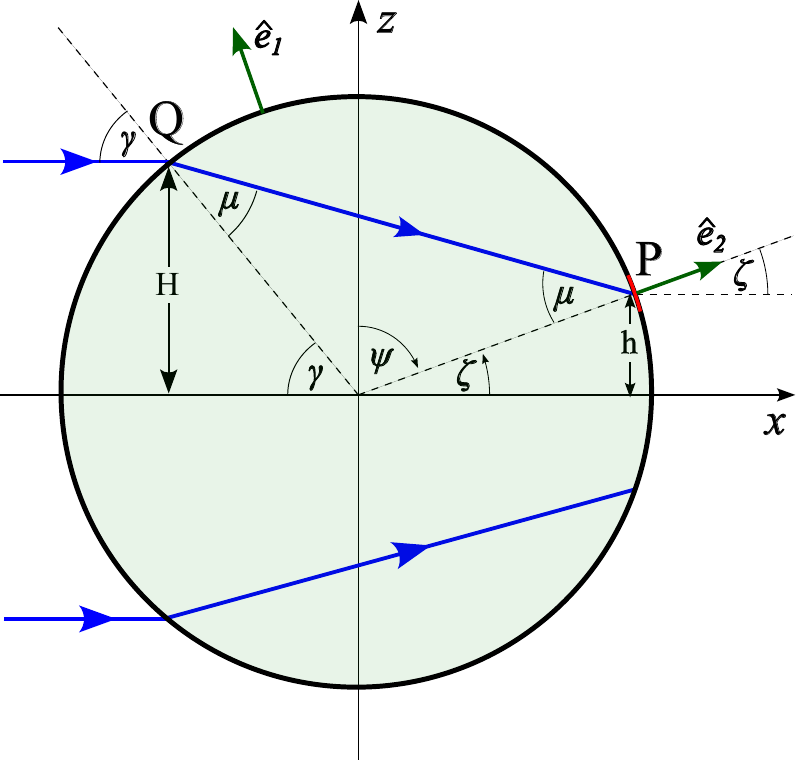}
    \caption{Cross section of a spherical cell illuminated from the left.  
    Cell swims with its body-fixed axis $\hat{\bf e}_3$ along the positive $y$-axis 
    and 
    rotates counterclockwise when viewed from behind, as in the drawing.  Visible body-fixed axes are 
    $\hat{\bf e}_1$ and $\hat{\bf e}_2$.
    The top ray of 
    light (blue) enters the cell at $Q$ and intercepts the photoreceptor (red arc) at $P$. 
    }
    \label{fig2}
\end{figure}

Consider the situation in which a cell swims toward $+y$ 
in the $xy$-plane while light of intensity $I_0$ shines toward $+x$ as in Figure \ref{fig2}, where the cell is
viewed from behind and rotates counterclockwise.  Choose the origin of coordinates to be the center $O$ of the cell
and let $\zeta\in [0,2\pi]$ be the angle of rotation around 
about $y$ (which is coincident with the direction $\hat{\bf e}_3$ of the body-fixed posterior-anterior 
axis), whose origin is chosen such that the angle indicating the location of the eyespot midpoint is also $\zeta$.  

For angles in the range $\pi/2\le \vert\zeta\vert \le \pi$ light is incident on the photoreceptor from the 
outside, with an average projection $I(\zeta)=\vert\cos\zeta\vert$, reaching a maximum when
the midpoint of the photoreceptor is on the $x$ axis.  

The quantity of interest now is the intensity $I(\zeta)$
for all other possible values of $\zeta$ when the light impinges on the photoreceptor from behind, 
having passed through the cell body. In previous work \cite{AlgalOptics}, we studied the 
high-symmetry case in
which the photoreceptor midpoint is again on the $x$ axis, but on the far side of the cell 
($\zeta=0$).  
In the limit $\epsilon \ll 1$, neglecting contributions from internal reflections, 
the on-axis intensity boost $\eta_0\equiv I(0)$ is \cite{AlgalOptics}
\begin{equation}
    \eta_0 = \left(\frac{n}{2-n}\right)^2.
    \label{boost}
\end{equation}
As mentioned above, for CR with $n=1.1$, $\eta_0$ has the surprisingly large value 
$\sim\! 1.5$.  (Note: $n=2$ corresponds to the lowest value at which the focal point of 
paraxial rays is inside the sphere.)

\begin{figure*}[t]
    \includegraphics[width=2.0\columnwidth]{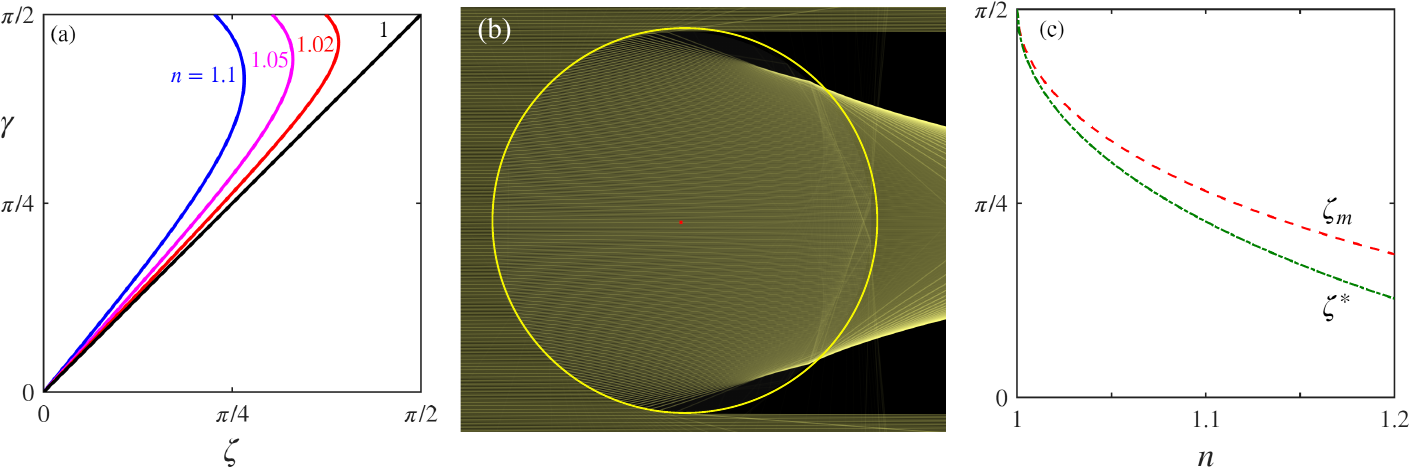}
    \caption{Geometrical optics for spherical cells. (a) Relationship \eqref{implicit} between the incident angle 
    $\gamma$ and the angle $\zeta$ of the 
    photoreceptor shown in Fig. \ref{fig2}. (b) Ray-tracings 
    \cite{rayoptics} illustrating caustic formation 
    for relative index of refraction $n=1.1$. (c) The angle 
    $\zeta_m$ of the caustic and angle $\zeta^*$ bounding the
    double-valued region of light, as function of the relative
    index $n$.}
    \label{fig3}
\end{figure*}

To find the intensity boost for all values of $\zeta$, we must find (see Fig. \ref{fig2}) 
which one of the incoming parallel lights rays refracts at point $Q$ on the cell boundary 
and impinges on the photoreceptor at $P$.  
The angles of incidence and refraction with respect to the surface normal at $Q$ are 
$\gamma$ and $\mu$, respectively, and they obey Snell's law,
\begin{equation}
    \sin \gamma = n \sin \mu.
    \label{snell}
\end{equation}
From elementary geometry we find 
\begin{equation}
    \zeta=2\mu-\gamma.
    \label{trig}
\end{equation}
Viewing $\zeta$ as the independent variable, 
Eqs. \eqref{snell} and \eqref{trig} provide two equations for the
two unknowns $\gamma(\zeta)$ and $\mu(\zeta)$ which are solved by the implicit
equations 
\begin{subequations}
\begin{align}
     2\sin^{-1}\left(\frac{\sin\gamma}{n}\right)-\gamma=\zeta, \label{implicit1}\\
     2\mu-\sin^{-1}\left(n\sin\mu\right)=\zeta.\label{implicit}
     \end{align}
      \label{implicit}
\end{subequations}
For $n=1$, when there is no lensing, $\gamma=\mu=\zeta$, as expected.
For small $\zeta$, one finds the expansions
\begin{equation}
    \gamma(\zeta)=\frac{n}{2-n}\zeta + \cdots, \ \ \ \ \mu(\zeta)=\frac{1}{2-n}\zeta + \cdots,
    \label{localboost}
\end{equation}
which are useful in determining the on-axis limit of the more general lensing expression below.

Figure \ref{fig3}(a) shows the function $\gamma(\zeta)$ 
for $n=1.1$, and illustrates that there is a
maximum possible value $\zeta_m$ for 
$\gamma\in [0,\pi/2]$.  The physical significance
of this is seen in Fig. \ref{fig3}(b), which displays
ray-tracings \cite{rayoptics} for a circular lens with
$n=1.1$.
We see that the maximum $\zeta$ is associated with 
the formation of caustics; no light reaches the 
photoreceptor for $\zeta_m < \vert\zeta\vert < \pi/2$.
We refer to this annular domain as the ``dark region".

The maximum angle $\gamma_m$ associated with 
$\zeta_m$ as a function of index 
of refraction ratio $n$ can be found by the 
condition $d\zeta/d\gamma=0$,
where
\begin{equation}
    \frac{d\zeta}{d\gamma}=\frac{2\cos(\gamma)}{\sqrt{n^2-\sin^2(\gamma)}}-1,
\end{equation}
yielding
\begin{equation}
    \gamma_m=\cos^{-1}\sqrt{\frac{n^2-1}{3}}.
\end{equation}

A second characteristic angle is associated with an incident beam that grazes the top 
of the circle, with incident angle $\gamma^*=\pi/2$, corresponding to
\begin{equation}
    \zeta^*=2\sin^{-1}\left(\frac{1}{n}\right)-\frac{\pi}{2}.
\end{equation}
For photoreceptor angles $\vert\zeta\vert<\zeta^*$ there is a unique incident ray 
that impinges on the 
photoreceptor, while for $\zeta^* < \vert\zeta\vert <\zeta_m,$ two distinct incident light rays, 
with two distinct angles $\gamma$, refract to the same point given by $\zeta$.  
As an example, for $n=1.1$ we find $\gamma_m\approx 74.5^\circ$, 
$\zeta_m\approx 47.8^\circ$ and $\zeta^*\approx 40.8^\circ$.  The dependence of 
$\zeta_m$ and $\zeta^*$ on $n$ is shown in Fig. \ref{fig3}(c).

\begin{figure}[b]
    \centering
    \includegraphics[width=\columnwidth]{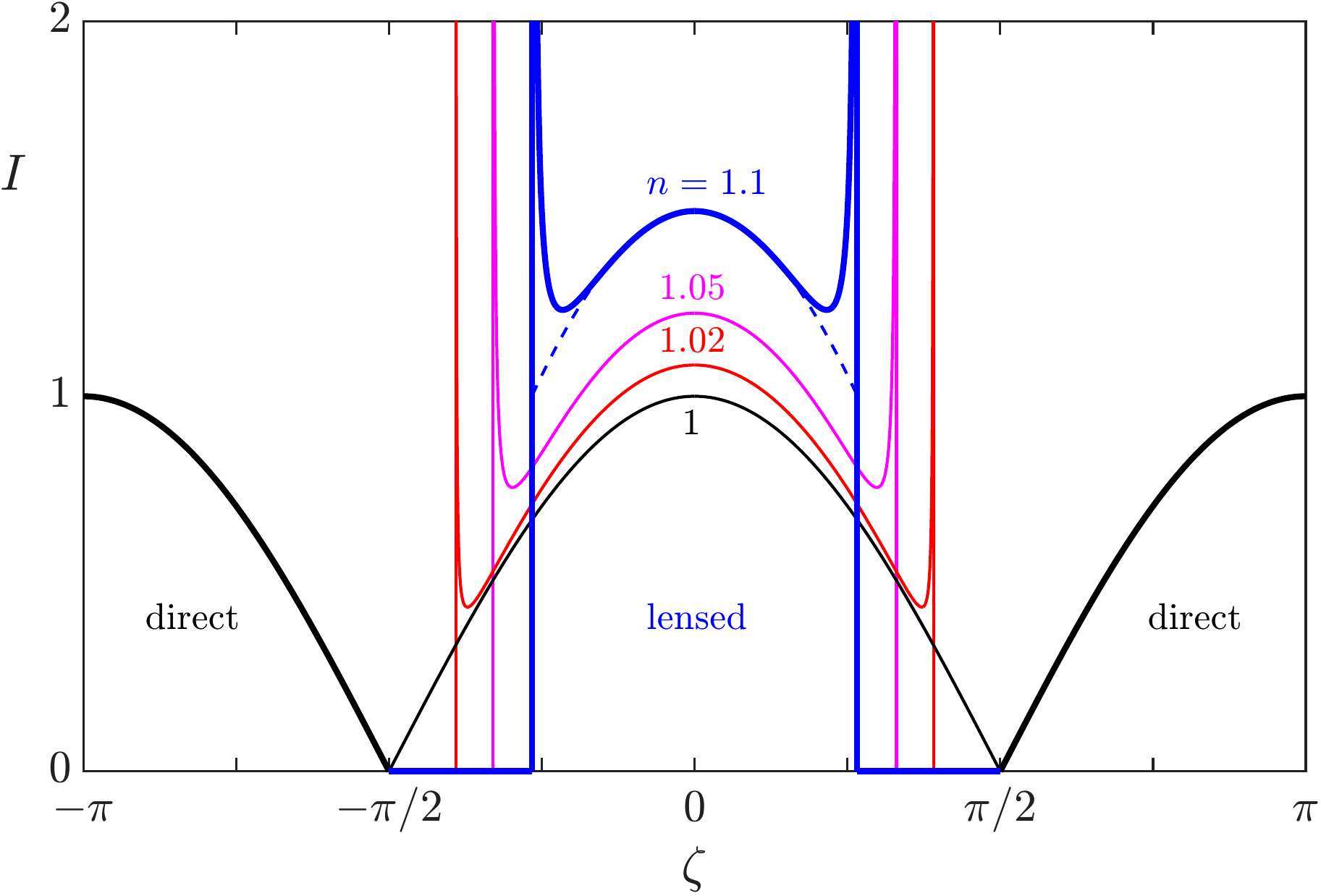}
    \caption{Angular dependence of lensing. Intensity versus rotation angle $\zeta$ for various values of $n$. Direct contributions are shown as heavy black curves.  Dashed blue line is local approximation 
    near $\zeta=0$ for representative value $n=1.1$.}
    \label{fig4}
\end{figure}

With these results, we may now compute the intensity
boost for a vanishingly small photoreceptor whose
midpoint is at angle $\zeta$.  We do this by 
calculating the ratio between the size of the 
incoming bundle of light rays that ultimately 
impinges on the photoreceptor to the size of 
the receptor itself. From 
Fig. \ref{fig2}, let $H=R\sin\gamma$ be the 
the distance from 
the $x$-axis to $Q$ and 
$h=R\sin\zeta$ be the distance from the $x$-axis 
to $P$.  Under a small 
change $\delta\gamma$, viewing $\zeta$ as a
function of $\gamma$, we find
\begin{equation}
    \delta H=R\cos\gamma\,\delta \gamma, \ \ \ \ 
    \delta h=R\cos\zeta \frac{d\zeta}{d\gamma}\delta\gamma.
\end{equation}

From trigonometry, we find that the bundle of rays 
heading from $Q$ to $P$ has a cross-sectional width  
$\delta h_\perp=\delta h\cos\mu/\cos\zeta$, from which it follows that
\begin{equation}
    \eta=\frac{2\pi H \delta H}{2\pi h\delta h_\perp}
    =\frac{\sin\gamma\cos\gamma}{\sin\zeta\cos\mu \, d\zeta/d\gamma},\label{etadefine}
\end{equation}
and the projection $I$ of
the refracted beam onto the photoreceptor is simply $I=\eta\cos\mu$ or
\begin{equation}
I(\zeta)=\frac{\sin\gamma\cos\gamma}{\sin\zeta \, d\zeta/d\gamma},
    \label{Ieta}
\end{equation}
where $\gamma(\zeta)$ is the solution to the implicit equation \eqref{implicit1}.
The result \eqref{Ieta} applies directly for $\vert\zeta\vert\in 
[0,\zeta^*]$ where the function $\gamma(\zeta)$ is single-valued.
In the interval $\vert\zeta\vert\in [\zeta^*,\zeta_{\rm m}]$ one
must add the separate contributions from the two branches of
$\gamma(\zeta)$.

Figure \ref{fig4} shows the complete angular dependence of
the light intensity, both for illumination and the lensed 
regions, for several values of $n$.  For any $n>1$ we see
that the intensity diverges at the angle $\zeta_m$, the 
location of the caustic, because the
derivative $d\zeta/d\gamma=0$ there.  But around $\zeta=0$ 
the lensed intensity is just a magnified form of the underlying
$\cos\zeta$ dependence; the dashed blue line in Fig. \ref{fig4}
is the function $\eta_0\cos\zeta$, with $\eta_0$ given by 
\eqref{boost}. Additionally, one verifies from \eqref{localboost} and 
\eqref{Ieta} that the on-axis boost result \eqref{boost} is
recovered as $\zeta\to 0$, as  
when $\gamma, \mu, \zeta\ll 1$, $\sin\gamma/\sin\zeta\approx \gamma/\zeta\approx d\gamma/d\zeta=n/(2-n)$, and $\cos\gamma\approx 1$, so $I(0)=(n/(2-n))^2$.

The cylindrical symmetry of the setup shown in Fig. \ref{fig2} implies that pattern of
lensed light within the cell is axisymmetric about $x$.  In \S \ref{adapt} we show how
this symmetry can be used to find the light intensity at the photoreceptor when 
it is out of the
$xz$-plane.

\section{Adaptive Phototaxis with Lensing}
\label{adapt}

We now embed the calculation of the previous section in a 
theory of \textit{Chlamydomonas} phototaxis \cite{ChlamyPRE},
illustrating the result in the context of the simplest geometry
of a phototurn: cells swimming in the $xy$-plane with 
a collimated source of light shining in the positive $x$ direction.  

\begin{figure}[t]
\includegraphics[width=0.6\columnwidth]{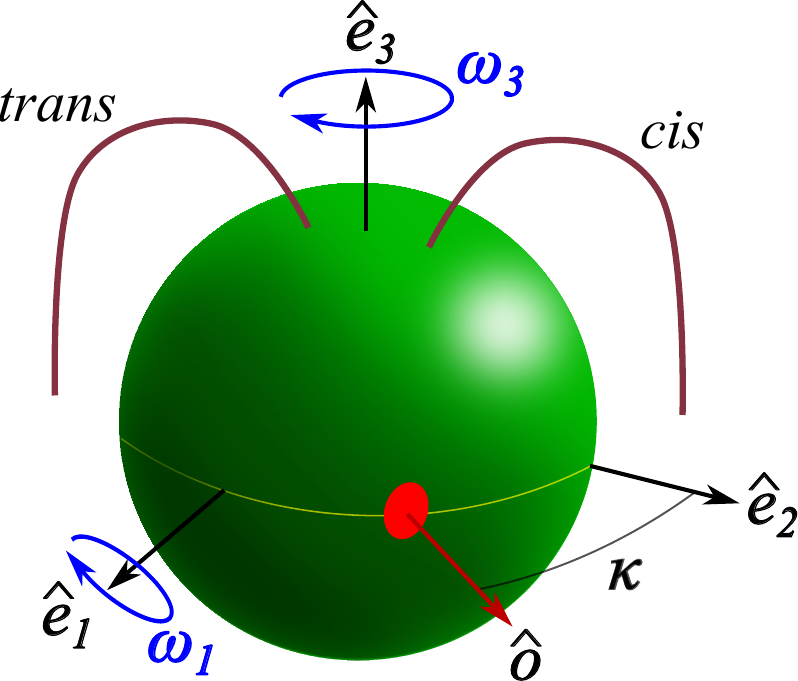}
\caption{Coordinate system of \textit{Chlamydomonas}, adapted
from \cite{Raikwar2025}.  Eyespot is shown as red disc, 
with outward normal $\hat{\bf o}$.  Cell rotates continuously 
around $\hat{\bf e}_3$ and transiently around $\hat{\bf e}_1$ 
due to the photoresponse of its \textit{cis} and \textit{trans} 
flagella.}
\label{fig5}
\end{figure}

In the model, a full account of swimming motions involves 
coupling the general rigid-body 
equations of motion for the three Euler angles $(\phi,\theta,\psi)$ 
with dynamics describing how the relevant angular velocities 
$(\omega_1,\omega_2,\omega_3)$ about the body-fixed axes 
$(\hat{\bf e}_1,\hat{\bf e}_2,\hat{\bf e}_3)$ evolve as light 
falls on the photoreceptor and the flagellar beating dynamics 
change accordingly.  With reference to Fig. \ref{fig5}, 
we choose $\hat{\bf e}_1$ to point normal to the plane 
containing the two flagella, $\hat{\bf e}_2$ to point towards 
the \textit{cis} flagellum (that which is closer to the 
eyespot/photoreceptor) and $\hat{\bf e}_3$ to point 
normal to the cell anterior, in the direction of swimming.
The eyespot is generally located close to the cell midplane, at some
angle $\kappa$ with respect to $\hat{\bf e}_2$
so that the eyespot normal is $\hat{\bf o}=\sin\kappa\, \hat{\bf e}_1 + \cos\kappa\,\hat{\bf e}_2$ 
\cite{ChlamyPRE}.  The results we obtain below are insensitive to the value of $\kappa$ and for
concreteness we set $\kappa=0$, so $\hat{\bf o}=\hat{\bf e}_2$, as in Fig. \ref{fig2}.

Because of the planarity of the flagellar beating, we may neglect
any rotations around $\hat{\bf e}_2$, thus setting 
$\omega_2=0$ and focus only on (i) steady spinning of the 
cell around $\hat{\bf{e}_3}$ at constant 
frequency $\omega_3<0$ and (ii) the changes
to $\omega_1$ arising from differences in the waveforms of
the \textit{cis} and \textit{trans} flagella induced by changing illumination
of the photoreceptor.
Ignoring small out-of-plane motions, we fix $\theta=\pi/2$ \cite{ChlamyPRE},  
and thus specialize to motions solely in the $xy$-plane, 
and find that the equations of motion for the Euler 
angles reduce to
\begin{subequations}
\begin{align}
    \dot{\phi}&=\omega_1\sin\psi, \label{phidot}\\
   \dot{\psi}&=\omega_3, \label{psidot}
\end{align}
\end{subequations}
where $\phi$ is the angle between $\hat{\bf e}_3$ and the negative $y$-axis and $\psi$ is the angle between 
$\hat{\bf e}_2$ and the $z$-axis.  The adaptive dynamics is
\begin{align}
    \tau_r\dot{\omega}_1&=s-h-\omega_1,\nonumber \\
    \tau_a\dot{h}&=s-h.
\end{align}
Here, $h$ is a hidden variable that accounts for adaptation, 
$\tau_r$ and $\tau_a$ are the (short) response and (long) 
adaptation time scales governing beating asymmetries, and 
$s=\omega_1^*I$ is the signal, governed by the light intensity 
$I$ at the
photoreceptor, parameterized in magnitude by the 
maximum angular frequency $\omega_1^*$
induced by the direct illumination.

We adopt the rescalings $T=\vert\omega_3\vert t$,
$P=\omega_1/\vert \omega_3\vert$, 
$P^*=\omega_1^*/\vert \omega_3\vert$,
$\alpha=\vert\omega_3\vert \tau_a$, $\beta=\vert\omega_3\vert \tau_r$, $H=h/\vert\omega_3\vert$
and $S=s/\vert\omega_3\vert$.  
Integrating $\psi_T=-1$ and considering the relationship $\zeta=\pi/2-\psi$ between the Euler angle $\psi$ and
the angle $\zeta$ in Fig. \ref{fig2}, we choose an integration constant in \eqref{psidot} for
later convenience and obtain $\zeta(T)=T$. As in previous work \cite{Raikwar2025}, we set 
$\phi=\pi/2+\varphi$, where $\varphi$ is the angle between the swimming direction and the positive 
$x$-axis.  The dynamics becomes
\begin{subequations}
\begin{align}
\varphi_T &= P\cos T,\label{fulldyna} \\
 \beta P_T&=S-H-P,\label{fulldynb}\\
 \alpha H_T&=S-H.\label{fulldync}
\end{align}
\label{fulldynamics}
\end{subequations}
When the light shines toward the positive $x$-axis ($\hat{\bf l}=\hat{\bf x}$),
and the normal vector to the photoreceptor is 
\begin{equation}
    \hat{\bf o}=\cos\zeta \left[ \sin\varphi\, \hat{\bf x}-\cos\varphi\, \hat{\bf y}\right]
    +\sin\zeta\,\hat{\bf z},
\end{equation}
the signal $S$ is related to the  
projection 
\begin{equation}
    J=-\hat{\bf l}\cdot \hat{\bf o}=-\sin\varphi\cos T
    \label{Jdef}
\end{equation}
through
\begin{equation}
    S=P^*\times\begin{cases}
        I_{\rm direct}, & \text{if} \ \  J\ge 0,\\
        I_{\rm lensed}, & \text{if} \ \ J<0,
    \end{cases}
    \label{signalcases}
\end{equation}
where $I_{\rm direct}=J$ and $I_{\rm lensed}$ is the result for arbitrary $\varphi$ of the calculation 
presented in \S \ref{lensing} for the special case $\varphi=\pi/2$.  To find the general expression, 
note that the axisymmetry of the lensed light implies that the intensity depends only on the 
angle $\zeta_{\rm eff}$ between the eyespot normal and the positive $x$-axis.
Simple geometry shows that 
\begin{equation}
    \zeta_{\rm eff}=\cos^{-1} \left(-J\right),
    \label{zetaeff}
\end{equation}
where $J$ is given by \eqref{Jdef}. It follows that $I_{\rm lensed}$ is given by \eqref{Ieta} with the substitutions $\zeta\to \zeta_{\rm eff}$
and $\gamma\to \gamma_{\rm eff}$, where $\zeta_{\rm eff}$ and $\gamma_{\rm eff}$ also satisfy 
\eqref{implicit1}.  In the limit $n\to 1$, where the lensing effects vanish and the light from behind 
the cell goes straight through, we have $\zeta_{\rm eff}=\gamma_{\rm eff}$ and thus $I\to \cos\zeta_{\rm eff}=\sin\varphi\cos T$, as
expected.

In this rescaled system of units the swimming trajectories ${\bf R}(T)$ are obtained by 
integrating the equations
\begin{equation}
    {\bf R}_T=U\left(\cos\varphi \hat{\bf x}+\sin\varphi \hat{\bf y}\right),
    \label{Reqn}
\end{equation}
where $U=u/(R\vert\omega_3\vert)$ is the swimming speed $u$ scaled by the cell radius $R$ and rotational
frequency.  The model is now completely defined by Eqs. \eqref{fulldynamics}, \eqref{Jdef}-\eqref{Reqn},
with parameters $\alpha, \beta, P^*, n$, and $U$.

\begin{figure*}[t]
\includegraphics[width=2.0\columnwidth]{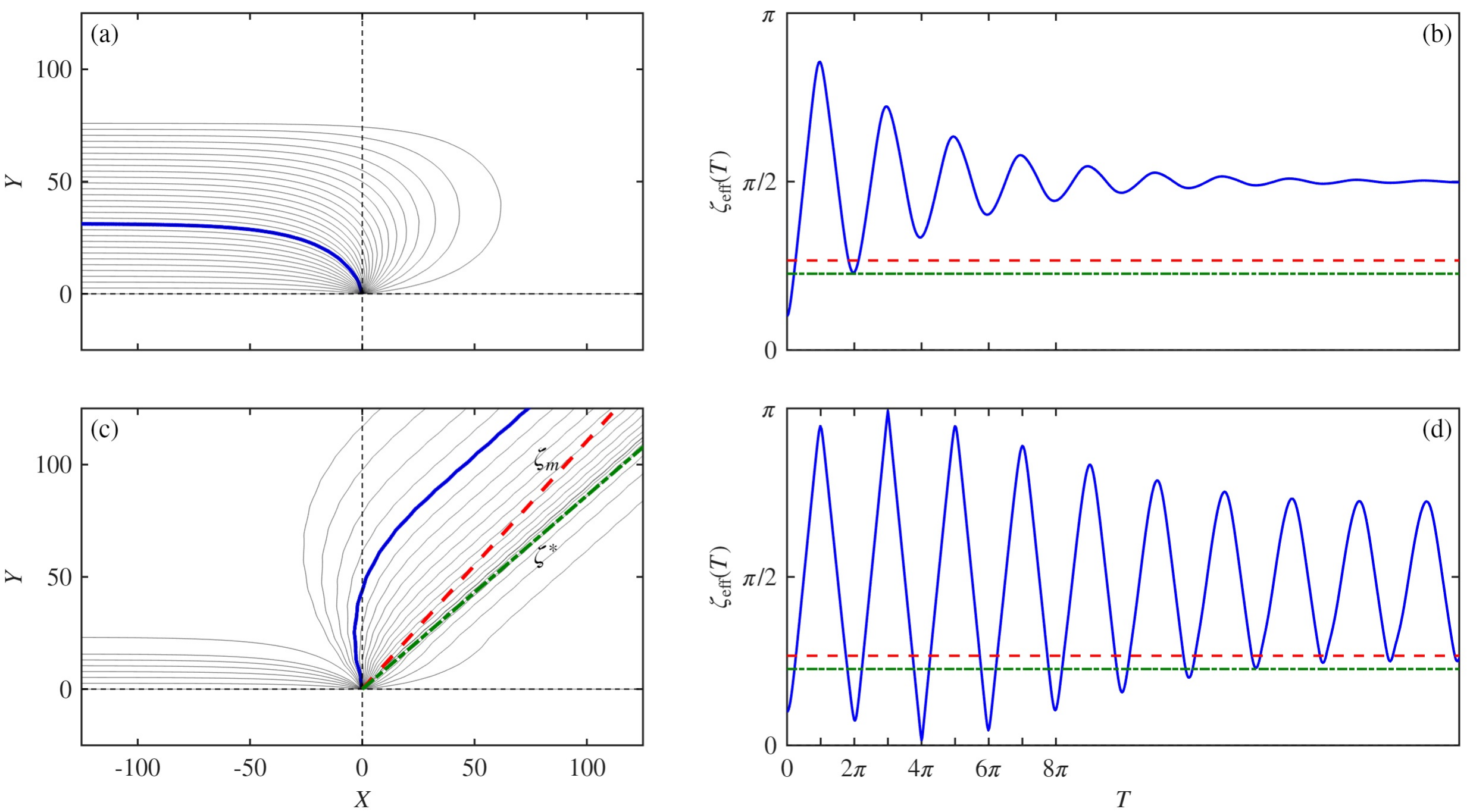}
\caption{Numerical results. (a) Trajectories in $XY$-plane for a range of initial orientations $\varphi\in [0,\pi]$
for a positively phototactic cell without lensing, with parameters
$\alpha=2, \beta=0.5, P^*=-0.4$ and $U=2$.  Trajectory for initial condition $\varphi(0)=0.55\pi$ is 
shown in blue and the associated projection angle $\zeta_{\rm eff}$ is shown in (b), where 
dashed lines indicate angles $\zeta_m$ and $\zeta^*$. All initial orientations 
lead to positive phototaxis.  (c) Trajectories as in (a) but with lensing.  A small subset of initial 
orientations lead to positive phototaxis, but the majority produce negative phototaxis. Blue 
trajectory in (c) has same initial orientation as in (a). (d) As in (b), but with lensing. Dashed lines in 
(c) are lines at
angles $\zeta_m$ and $\zeta^*$.}
\label{fig6}
\end{figure*}

We now turn to numerical results on phototaxis in the presence of lensing, show that lensing
indeed produces a reversal in the sign of phototaxis, and propose a physical explanation for the effect.
We continue with the geometry assumed in the previous section: a cell swimming in the $xy$-plane with
light shining along the positive $x$-axis.  Figure \ref{fig6} shows (dark blue) the trajectory of a cell which starts
at the origin with an 
initial orientation angle $\varphi=0.55\times \pi$, so that it faces slightly toward the light.
In Fig. \ref{fig6}(a) we consider the wild-type cell with eyespot shading and hence no lensing.
The cell rapidly turns toward the light and ultimately swims toward it directly along the negative
$x$-axis ($\varphi=\pi$).  Figure \ref{fig6}(b) shows the projected angle $\zeta_{\rm eff}$ versus time,
illustrating that during the first few cellular rotations some lensed light would have been experienced 
were it not for eyespot shading, but eventually $\zeta_{\rm eff}\to \pi/2$ as the cell aligns with the light
and the eyespot is orthogonal to the light.  
The gray trajectories shown in Fig. \ref{fig6}(a),
for initial angles uniformly spaced 
within the interval $[0,\pi]$, illustrate that the 
positive phototaxis holds for all initial conditions.

In contrast to the wild-type behavior, Fig. \ref{fig6}(c) shows the trajectory (dark blue) for the 
same initial condition when lensed light is 
sensed.  Although initially heading slightly towards the light, the cell turns around and swims 
away from it, exhibiting negative phototaxis and settling into a straight trajectory whose angle $\varphi^*$ is between the two angles
$\zeta^*$ and $\zeta_m$ associated with the lensing effect.  This is seen clearly in 
Fig. \ref{fig6}(d), where, during each rotation of the cell, $\zeta_{\rm eff}$ reaches a 
minimum between $\zeta^*$ and $\zeta_m$ and a maximum that is well beyond $\pi/2$.  Thus, in 
the steady state swimming at long times, the photosensor detects two signals during each cellular
rotation.
These
two signals induce phototaxis in opposite directions and those two effects balance to yield
the compromise direction $\varphi^*$, whose precise value depends on the position of the photoreceptor relative to the
cell midplane \cite{Raikwar2025} and the relative index $n$.  As in 
Fig. \ref{fig6}(a), trajectories for a range of initial
orientations are shown in gray, and indicate that there is
a small range of initial angles near $\pi$ which lead
to positively phototactic swimming.  The great majority
of initial conditions leads to negative phototaxis.

In order to understand these results, we 
note that the chief difference between the incident and 
lensed profiles is that the latter is confined to a
smaller angular domain (see Figs. \ref{fig3} and \ref{fig4}) and not only has a larger amplitude 
within that domain, but also has much higher derivatives due to the presence of the caustic at the
domain edges.   This difference holds in spite of the fact that 
the integrated flux of light incident
on the cell is identical to the integrated lensed flux incident on the interior wall of the cell.
We thus conclude that the reverse in sign of phototaxis is the consequence of the dominance of the 
lensed photoresponse over the direct one due to the larger time derivative of the lensed signal, 
as explained below.

That it is the time derivative of the signal that determines the flagellar photoresponse can be 
traced back to the pioneering work of R{\"u}ffer and Nultsch \cite{RufferNultsch1991} who found that
the \textit{cis} flagellum increases (decreases) its beating amplitude during a step up (down) in
light intensity, while the \textit{trans} flagellum does the opposite.  This asymmetry is necessary
for wild-type phototaxis, as evidenced by the mutant strain \textit{ptx1} which lacks this asymmetry
and does not do proper phototaxis \cite{RufferNultsch1998}.   The role of this asymmetry in phototaxis 
can be seen by considering the photoresponse as the photoreceptor rotates first toward the light and then
away, so the derivative $S_T$ changes sign.  The changes in flagellar beating when $S_T>0$ must be 
opposite those when $S_T<0$ half a turn later, or else the two responses will cancel and there will be
no net turn toward the light.  

The observation of R{\"u}ffer and Nultsch is actually embodied in the adaptive model of flagellar dynamics, 
even though the dynamics in \eqref{fulldynb} and \eqref{fulldync} are forced by the signal $S$ itself. 
As shown elsewhere \cite{Raikwar2025}, these coupled equations for $P$ and $H$ 
can be recast as a single equation for the photoresponse
variable $P$, 
\begin{equation}
P_{TT}+\frac{\alpha+\beta}{\alpha\beta}P_T+\frac{1}{\alpha\beta}P=\frac{1}{\beta}S_T,
\label{system1b}
\end{equation}
which is that of a damped harmonic oscillator, forced by the time derivative $S_T$ of the
signal.   Recalling that $\alpha=\vert \omega_3\vert \tau_a=2\pi \tau_a/t_{rot}$, where $t_{rot}$ is the 
rotation period, we deduce that for $\tau_a/t_{rot}< 1$ ($\alpha < 2\pi$) there is sufficient time
during a half-rotation for the adaptive response to reset itself and produce an antisymmetric response
to the changing light levels, and the smaller the value of $\alpha$ the more accurately antisymmetric
is the response and the more $P$ mirrors the time derivative of the signal.  

Moreover, the origin of the trajectories in 
Fig. \ref{fig6}(c)
that head toward the light can also be understood 
as a consequence of the confinement of the lensed light to a small
angular domain; the photoreceptor of a 
cell that swims 
in a direction $\varphi$ significantly larger than 
$\pi/2$ will receive direct illumination during one
half of the rotational cycle and will be in the dark
region during the other half.  This phenomenon is
thus analogous to the situation described recently 
\cite{Raikwar2025} when \textit{Chlamydomonas} is 
illuminated by two nearly antiparallel sources.  When
the eyespot is displaced from the cell midplane, there 
is a range of swimming directions within which only a 
single source is sensed.

We have presented numerical results for the case in which a wild type cell exhibits positive phototaxis, 
corresponding to the choice $P^*<0$ \cite{ChlamyPRE}, and lensing effects induce negative phototaxis. 
The symmetry of the problem is such that to describe a negatively phototactic wild type cell requires 
only changing the sign of $P^*$, and lensing effects would then induce positive phototaxis.

In order to interpret the experimental results of Ueki, et al. \cite{Ueki2016}, 
where cell accumulation at the edge of a quasi-two-dimensional algal suspension in a Petri dish was 
used as a measure of phototactic movement, we apply the two-dimensional theoretical results discussed above.
These imply that for a given initial position and orientation within the Petri dish, 
positively phototactic 
cells subjected to a collimated beam of light would swim away from the light 
asymptotically along one of two rays 
at $\pm\varphi^*$, and they would intersect the edge of a petri dish at one of two spots separated by
the angle $2\varphi^*$. Averaging over initial positions and orientations will produce a continuous 
distribution of accumulation at the Petri dish edge, as seen in experiments.  
This smearing effect will be enhanced by at least four effects: (i) the distribution of photoreceptor
positions relative to the cell equator \cite{Raikwar2025}, (ii) the finite size of the photoreceptor 
(which will smear out the caustic), and (iii) the varying photoreceptor tilts relative to the 
swimming direction due to a distribution of helical trajectories, and
(iv) the inherent noisiness of the trajectories due to the 
``run-and-turn" locomotion \cite{Polin2009} 
associated with stochastic switching between 
periods of flagellar synchrony and asynchrony.

\section{Discussion}
\label{disc}

In this paper we have provided a quantitative theory for the 
experimentally observed sign reversal of phototaxis in mutants of
the unicellular green alga \textit{Chlamydomonas} that do not have
the pigmented eyespot responsible for directional sensitive of the
photoreceptor.  Our results indicate that the sign reversal arises 
not from the brightness of the lensed light itself, but rather from 
the greater rate of change of the intensity of lensed light experienced by the 
photoreceptor as a cell turns. The analysis predicts that reversed-sign phototaxis
is associated with swimming away from the light at a particular angle, 
and that a population of positively phototactic cells with randomized initial positions and orientations would
display bulk movement away from the light within a broad angular distribution, as observed 
experimentally, but with a subset of
the population moving toward the light. 
While existing studies \cite{Ueki2016} of the relevant phototactic mutants have not studied 
single-cell swimming with sufficient detail to quantify individual trajectories and test these
predictions, 
2D and 3D tracking methods
of the kind that have been developed for studying protists 
\cite{ChlamyPRE,GoniumPRE,Drescher2009} can be used.

\begin{acknowledgments}
This work was supported 
in part by a Trinity College Summer Studentship and 
a Gates Scholarship (M.Y.),
Gordon and Betty Moore Foundation Grant No. 7523 (S.K.B. \& R.E.G.) 
and Wellcome Discovery Award 307079/Z/23/Z (A.I.P. \& R.E.G.).
\end{acknowledgments}

\end{document}